\def\fun#1#2{\lower3.6pt\vbox{\baselineskip0pt\lineskip.9pt
\ialign{$\mathsurround=0pt#1\hfil##\hfil$\crcr#2\crcr\sim\crcr}}}
\relax
\documentstyle[12pt]{article}

\advance\textheight by \topskip
\begin{document}
\begin{flushright}
DAMTP R94-62\\
hep-th/9505147
\end{flushright}
\vspace{-0.2cm}
\begin{center}
{\Large\bf Structure of dualities in bosonic string theory}\\
\vskip 1.7 cm
{\bf
Ian R. Pinkstone  }
\footnote{E-mail: I.R.Pinkstone@amtp.cam.ac.uk}
\vskip 1cm
{\em Department of Applied Mathematics and Theoretical Physics\\
University of Cambridge\\
Silver Street\\
Cambridge\\
CB3 9EW\\
England\\}
\vskip 0.7 cm
\end{center}
\vskip 1.5 cm
\centerline{\bf ABSTRACT}
\begin{quotation}
The nature of duality symmetries is explored in closed bosonic string
theory, particularly in the case of a four-dimensional target space
admitting a
one-parameter isometry. It appears that the S-duality of string theory
behaves analogously
to the Ehlers' symmetry of General Relativity. Furthermore, it is
demonstrated
that the $O(1,1)$ target space duality arising from the isometry
interchanges the
roles of these two symmetries. The inclusion of the tachyon field is
shown to be
consistent with T-duality but incompatible with S-duality. Finally,
extrapolating
to dimensions other than four, the effective action is found to be
invariant under
a larger group of symmetries than the usual $O(1,1)$.
\end{quotation}
\newpage
{\center \section{Introduction}}
It is becoming increasingly apparent that string theory possesses a
far richer structure than is evident from the conventional
perturbative formulation,
containing many more symmetries than the
Standard Model and General Relativity which it will hopefully unite or
replace. Two
of these symmetries have become known as string dualities, explicitly
S-duality
and T-duality (target space duality). The former is an $SL(2, R)$
symmetry
acting on the equations of motion of the effective four dimensional
action of
heterotic string theory, which is related to the electric magnetic
duality of four dimensional field theory and may lead to a realization
of the strong
weak duality conjecture of Montonen and Olive \cite{MO}, under which
the weak
coupling limit exchanges with the strong coupling limit.  If exact,
this symmetry would
be of immense benefit: from a knowledge of low energy string theory
the structure of
strongly coupled string theory could be determined, a domain otherwise
only accessible
with as yet unknown non-perturbative techniques. Unfortunately, it is
this
non-perturbative behaviour that removes the possibility of proving
exactness
order by order in perturbation theory. However, there is some evidence
that
suggests that the duality is exact \cite{Sen}. As we shall see,
S-duality is also related in certain circumstances to the Ehlers'
symmetry of
General Relativity. T-duality, however, is an entirely new phenomenon,
unique to
string theory. It is most clearly demonstrated in the simple model of
a string moving
on a background consisting of a compact dimension of radius $R$: the
theory is
invariant under the duality transformation $R\rightarrow {1\over {2R}}$
(in units such
that $\alpha'=1$). This can be physically understood as an exchange of
the string momentum and winding modes. In the case of a target space
with $d$
compactified dimensions, this extends to an $O(d,d; Z)$ transformation
between two differing sets of target space fields, which preserves the
underlying conformal field theory. Unlike S-duality, T-duality has
been shown to beexact \cite{gpr}.

So we have: T-duality, which is a perturbatively exact symmetry,
acting in the
presence of compact target space dimensions; and
S-duality, conjectured to be exact, acting in an effective four
dimensional target
space. It would be interesting to investigate the combined effects of
S and T-duality. In \cite{bakas1}, S and T-duality were studied in
bosonic string theory
with two Abelian isometries, where it was shown that a string
generalization of the
Geroch group is obtained. We shall again look at bosonic string
theory, but concentrate on a thorough analysis of the case with
just one isometry and shall discover an interesting
phenomenon not immediately apparent in the study with two isometries.
At the level of this investigation,
i.e. looking only at the effective action, it does not matter whether
this isometry is compact or not, and it will be assumed that the
basic symmetries will be
$O(1,1; R)$ and $SL(2, R)$ for T-duality and S-duality respectively.

The effective action of the target space, $\cal{M}$, is
\begin{equation}
I=\int_{\cal{M}} d^4 \! x \sqrt{\tilde{g}} e^{\Phi}
\left(\tilde{R}(\tilde{g}) +
\left(\nabla \Phi \right)^2 - {1 \over 12} H_{\mu\nu\rho}
H^{\mu\nu\rho} \right) \ , \label{acs4}
\end{equation}
where $\tilde{g}$ is the target space metric, $\tilde{R}(\tilde{g})$
is the corresponding Ricci scalar, $\Phi$ is the dilaton, and
$H_{\mu\nu\rho}$ is
the potential of the Kalb-Ramond antisymmetric tensor $B_{\mu\nu}$,
\begin{equation}
H_{\mu\nu\rho} = \partial_\mu B_{\nu\rho} + \partial_\nu B_{\rho\mu} +
\partial_\rho B_{\mu\nu} \ . \label{kalb}
\end{equation}
The tachyon field has been set to zero for this analysis, but we will
study the effect of its inclusion in a later section.
Because the target space considered here is not of the critical
dimension
($d=26$ for the bosonic theory), the total central charge must be made
up to
zero. Including in the action a cosmological constant type term, as in
\cite{beta}, will be shown later to be incompatible with S-duality of
the
effective action. Instead,
we will assume that the effective four dimensional theory is tensored
with an
appropriate conformal field theory of the required central charge,
that can be
ignored for the purposes of this analysis. Alternatively, (\ref{acs4})
can be
considered as a truncation of critical bosonic or super/heterotic
string theory,
compactified on a suitable manifold.
This action actually represents the maximal sector that is common to
all of
the standard string theories: bosonic, superstring types I, IIA, IIB;
and heterotic.
The metric $\tilde{g}$ will be taken positive
definite, though the entire analysis goes through for any choice of
signature.
By the conformal rescaling  $\tilde{g}_{\mu\nu} = e^{-\Phi}
g_{\mu\nu}$, (\ref{acs4}) can be rewritten in the Einstein frame
\begin{equation}
I=\int_{\cal{M}} d^4 \! x \sqrt{g} \left(R(g) - {1 \over 2}
\left(\nabla \Phi \right)^2 - {1 \over 12} e^{2 \Phi} H_{\mu\nu\rho}
H^{\mu\nu\rho} \right) \ , \label{ace4}
\end{equation}
where all indices are now raised and lowered with ${g}$. For the
existence of
the isometry, there can be no boundaries in ${\cal{M}}$ normal to the
isometry.
It will be further assumed that ${\cal{M}}$ is compact, though this
condition may be relaxed with a proper treatment of surface terms.

\noindent Note that the following non-standard normalization will be
used for
symmetrization and antisymmetrization of tensors:
\begin{equation}
A_{(\mu_1...\mu_n)} = {1 \over {(n-1)!}}\left(A_{\mu_1...\mu_n} +
{\rm cyclic\
perms} + {\rm anticyclic\ perms} \right) \
\end{equation}
\begin{equation}
A_{[\mu_1...\mu_n]} = {1 \over {(n-1)!}}\left(A_{\mu_1...\mu_n} +
{\rm cyclic\
perms} - {\rm anticyclic\ perms} \right) \ .
\end{equation}
This particular normalization enables the following definitions to
take the form
\begin{equation}
F_{\mu\nu} = \partial_{[\mu} A_{\nu]} = \partial_\mu A_\nu -
\partial_\nu A_\mu \ ,
\end{equation}
\begin{equation}
H_{\mu\nu\rho} = \partial_{[\mu} B_{\nu\rho]} = \partial_\mu
B_{\nu\rho} +
\partial_\nu B_{\rho\mu} + \partial_\rho B_{\mu\nu} \ .
\end{equation}
Also, $\epsilon_{\mu\nu\rho\sigma}$ and $\epsilon_{ijk}$, represent
the totally
antisymmetric tensors in four and three dimensions respectively.

{\center \section{\bf S-Duality of the effective action}}

A simplified version of S-duality exists for bosonic string theory in
an effective four-dimensional background; it is simplified as there
are no Maxwell gauge fields, and hence the electric-magnetic duality
is absent. The remaining symmetry
acts upon the fields appearing in the action (\ref{acs4}). It is
interesting to note that the effect of S-duality decouples from the
metric field
precisely when the metric is conformally rescaled to the Einstein
frame. The
importance of this point will be seen later. For now this observation
simply leads
us to work with the Einstein frame action (\ref{ace4}) for ease of
computations.

There appears to be some confusion in the literature with regards to
the method
for constructing the manifestly S-duality invariant action from
(\ref{ace4}).
The 3-form, $H_{\mu\nu\rho}$, is usually dualized by writing
\begin{equation}
H_{\mu\nu\rho} = e^{-2\Phi}{\epsilon_{\mu\nu\rho}}^\sigma\partial_
\sigma\Psi \label{dH}
\end{equation}
where $\epsilon_{\mu\nu\rho\sigma}$ is the totally antisymmetric
tensor and $\Psi$ is the axion pseudo-scalar field. Substituting
(\ref{dH}) into the action gives
\begin{equation}
I=\int_{\cal{M}} d^4 \! x \sqrt{g} \left(R(g) - {1\over2}\left(\nabla
\Phi
\right)^2-{1\over2}e^{-2\Phi} \left(\nabla\Psi\right)^2 \right) \ .
\label{adH}
\end{equation}
However, the equations of motion derived from (\ref{adH}) do not agree
with those
from (\ref{ace4}). To obtain an action consistent with (\ref{ace4}) we
must
turn to the language of forms. Let $H_3$ be the three form
$H_{\mu\nu\rho}$, $H_1$ its Hodge dual, and
$B_2$ the two-form $B_{\mu\nu}$. Now $H_3=dB_2$ so $dH_3=0$, and
$H_3=*H_1$
so $d\!*\!\!H_1=0$, but not $dH_1=0$ which is essentially the
assumption
made in (\ref{dH}). Finally if $d\!*\!\!H_1=0$ then $*d\!*\!\!H_1=0$.
This last relation can be written in tensor notation as
\begin{equation}
\epsilon^{\mu\nu\rho\sigma}\nabla_{\!\mu}
{\epsilon_{\nu\rho\sigma}}^\tau H_\tau = 0 \ , \label{tensor}
\end{equation}
where $H_\tau$ is the Hodge dual of $H_{\mu\nu\rho}$, given by
\begin{equation}
H_{\mu\nu\rho} ={\epsilon_{\mu\nu\rho}}^{\tau} H_{\tau} \ .
\label{hogH}
\end{equation}
Equation (\ref{tensor}) simplifies to
\begin{equation}
\nabla_{\!\mu} H^{\mu} = 0 \ . \label{consH}
\end{equation}
The procedure is now clear: we substitute
(\ref{hogH}) into (\ref{ace4}) and add the constraint (\ref{consH})
multiplied
by a Lagrange multiplier $\Psi$
\begin{equation}
I=\int_{\cal{M}} d^4 \! x \sqrt{g} \left(R(g) - {1 \over 2}
\left(\nabla \Phi \right)^2 - {1 \over 2} e^{2 \Phi} H_{\!\mu} H^{\mu}
+ \Psi \nabla_{\!\mu} H^{\mu} \right) \ . \label{acl4}
\end{equation}
The equation of motion for $H_{\mu}$ is
\begin{equation}
e^{2 \Phi} H_{\!\mu} = - \nabla_{\!\mu} \Psi \ . \label{eomH}
\end{equation}
After inserting (\ref{eomH}), integrating by parts, and removing the
surface
term (\ref{acl4}) becomes
\begin{equation}
I=\int_{\cal{M}} d^4 \! x \sqrt{g} \left(R(g) - {1 \over
2}  \left( \left(\nabla \Phi \right)^2 - e^{- 2 \Phi} \left(\nabla
\Psi \right)^2 \right) \right) \ , \label{aca4}
\end{equation}
where $\Psi$ is now the axion pseudo-scalar field. There is an obvious
sign difference in the axion term between
(\ref{adH}) and (\ref{aca4}) and
the  latter does indeed lead to the same equations of motion as the
original action (\ref{ace4}). The dilaton-axion
term forms the Lagrangian of an $O(2,1)$ non-linear $\sigma$ model
which is invariant under three transformations:

\noindent \quad dilations
\begin{eqnarray}
\Phi & \rightarrow & \Phi + a \nonumber \\
\Psi & \rightarrow & e^a\Psi \ , \label{dilat}
\end{eqnarray}
\quad translations
\begin{eqnarray}
\Phi & \rightarrow & \Phi \nonumber \\
\Psi & \rightarrow & \Psi + b \ , \label{trans}
\end{eqnarray}
\quad and a non-trivial mixing
\begin{eqnarray}
e^{\Phi} & \rightarrow & {{e^{\Phi}}\over{(1-c\Psi)^2-c^2e^{2\Phi}}}
\nonumber
\\ \Psi & \rightarrow & {{\Psi+c(e^{2\Phi}-\Psi^2)}\over{(1-c\Psi)^2-
c^2\Psi^2}} \ . \label{mix}
\end{eqnarray}
These are more clearly shown by introducing two new scalar fields
\begin{equation}
\lambda_{\pm} = \Psi \pm e^{\Phi}
\end{equation}
and writing (\ref{aca4}) as
\begin{equation}
I=\int_{\cal{M}} d^4 \! x \sqrt{g} \left(R(g) -{1 \over 2}
{{\partial \lambda_{+} \partial \lambda_{-}} \over
{\left(\lambda_{+}\!-\! \lambda_{-} \right)^2}} \right) \ . \label{ac4}
\end{equation}
The transformations (\ref{dilat})-(\ref{mix}) can now be recognized as
an $SL(2, R)$ group of isometries realized by
\begin{equation}
\lambda_{\pm} \rightarrow \lambda_{\pm}^{\prime} = {{a \lambda_{\pm}\!
+ b} \over {c \lambda_{\pm}\! + d}} \ ,
\hspace{30pt} ad - bc = 1 \ . \label{sdual}
\end{equation}
Obviously (\ref{ac4}) is invariant under (\ref{sdual}) which
forms the S-duality of bosonic string theory.  For
the heterotic string, S-duality is reduced to a symmetry of the
equations
of motion of the effective action, and furthermore, it is believed
that the $SL(2, R)$ symmetry will be broken down to an $SL(2, Z)$
symmetry by instanton effects \cite{Sen}.

{\center\section{\bf S-duality in the presence of an isometry}}

Anticipating our study of T-duality, it would be useful to examine the
behaviour of S-duality when ${\cal M}$ contains a one-parameter
isometry generated by the Killing vector
$K = K^\mu \partial_\mu = \partial_0$. If
the metric were Lorentzian rather than Riemannian, $K$ could be either
time-like or space-like, but not null for the following analysis. Thus
\begin{equation}
{\cal{L}}_K(g_{\mu\nu}) = 0, \hspace{30pt} {\cal{L}}_K(B_{\mu\nu}) =
`0, \hspace{30pt} {\cal{L}}_K \Phi = 0 \ , \label{lie}
\end{equation}
where $\cal{L}$ is the Lie derivative.  The isometry defines a
fibering $\pi:{\cal M}-{\cal C} \rightarrow \Sigma$ where $\Sigma$ is
the three
dimensional base manifold and ${\cal C}$ is the fixed point set of the
isometry. Locally, the metric on ${\cal M}$ can be written \cite{gary}
\begin{equation}
g_{\mu\nu}=\left (\begin{array}{lc} e^U & e^U\omega_j \\[9pt]
e^U\omega_i &
e^U\omega_i \omega_j + e^{-U}\gamma_{ij} \end{array} \right) \ ,
\label{met}
\end{equation}

\begin{equation}
g^{\mu\nu}=\left (\begin{array}{lc} e^U\omega_i\omega_j \gamma^{ij}+
e^{-U} & -e^U\omega_i \gamma^{ij} \\[9pt] -e^U\omega_j \gamma^{ij}&
e^U \gamma^{ij} \end{array} \right) \ , \label{Met}
\end{equation}

\noindent where $e^{-U}\gamma_{ij}$ is the $d$-dimensional metric on
$\Sigma$, $U$ is a scalar field, and $\omega_i$ is
a gauge potential defined up to
\begin{equation}
\omega_i \rightarrow \omega_i + \partial_i \chi \ .
\end{equation}
The natural projection operator from
$\cal M$ down onto $\Sigma$ is
\begin{equation}
h^\mu_{\ \nu}=\delta^\mu_{\ \nu}-{K^\mu K_\nu \over \|K\|^2} \ ,
\label{proj}
\end{equation}
(note that if $g$ were Lorentzian, and $K$ null, (\ref{proj}) would be
ill-defined). A gauge invariant field on $\Sigma$ can be defined as
\begin{equation}
V_{ij}=\partial_i\omega_j-\partial_j\omega_i \ .
\end{equation}
The fields $V_{ij}$ and $\gamma_{ij}$ are tensors on $\Sigma$ and
their indices
are raised and lowered with $\gamma$. Using (\ref{met}) and (\ref{Met})
the curvature tensors on $\cal M$ can be expressed in terms of $U$,
$V_{ij}$,
$\gamma_{ij}$, and their derivatives. In particular, the Ricci scalar
becomes
\begin{equation}
R(g)=e^UR(\gamma) + e^U D^2U - {1 \over 2} e^U(DU)^2 -
{1 \over 4} e^{3U}V_{ij}V^{ij} \ , \label{R}
\end{equation}
where R($\gamma$) is the Ricci scalar of the metric $\gamma_{ij}$ on
$\Sigma$, and $D_i$ is the covariant derivative induced
by $\gamma_{ij}$ on $\Sigma$. All indices are raised and
lowered with $\gamma$. The determinant of $g_{\mu\nu}$ is
\begin{equation}
\det (g_{\mu\nu}) = e^{-2U} \det (\gamma_{ij}) \ . \label{det}
\end{equation}
By (\ref{lie}), the dilaton term in (\ref{aca4}) can be simply
expressed as
\begin{equation}
g^{\mu\nu}\nabla_{\!\mu} \Phi\nabla_{\!\nu}\Phi = e^U \gamma^{ij}
D_i\Phi D_j\Phi\ . \label{dilred}
\end{equation}
However, the axion needs more careful treatment. From (\ref{lie})
\begin{equation}
\partial_0 B_{\mu\nu} = 0, \quad\quad \partial_0 H_{\mu\nu\rho} = 0 \ ,
\end{equation}
from which it follows, using (\ref{hogH}), that $\partial_0 H_\mu = 0$.
This implies by (\ref{eomH}) that
\begin{equation}
\partial_0 \partial_\mu \Psi = 0 \ , \label{axderiv}
\end{equation}
but not $\partial_0 \Psi = 0$. $\nabla_{\!\mu} \Psi$ can be projected
down from $\cal M$ onto $\Sigma$ using the projection
operator (\ref{proj})
\begin{eqnarray}
\partial_0\Psi=&K^\mu\nabla_{\!\mu}\Psi&=\ \nabla_{\!0}\Psi\nonumber \\
D_i\Psi=&h^\mu _{\ i}\nabla_{\!\mu}\Psi & = \
\nabla_{\!i}\Psi-\omega_i \nabla_{\!0}\Psi \ .
\end{eqnarray}
$\nabla_{\!\mu} \Psi$ can now be written as
\begin{equation}
\nabla_{\!\mu} \Psi=\left (\begin{array}{c} \partial_0\Psi \\[9pt]
D_i\Psi + \omega_i \partial_0\Psi \end{array} \right) \ . \label{ax3}
\end{equation}
Thus
\begin{equation}
g^{\mu\nu} \nabla_{\!\mu} \Psi \nabla_{\!\nu} \Psi = e^U \gamma^{ij}
D_i\Psi D_j\Psi + e^{-U} (\partial_0\Psi)^2 \ . \label{axred}
\end{equation}

\noindent We can substitute (\ref{R}), (\ref{det}), (\ref{dilred}),
and (\ref{axred}) into (\ref{aca4}), integrate the
$D^2U$ term by parts, and remove the surface term to obtain
\begin{equation}
I=\int_{\cal{M}}\! d^4 \! x \sqrt{\gamma} \left(\!R(\gamma)\! -\!
{1 \over 2} \!\left(\!DU \right)^2 \!- \!{1 \over 4} e^{2U} V_{ij}
V^{ij}\!-\!{1 \over 2} \!\left(\left(\!D\Phi\!\right)^2\!-\!
e^{-2\Phi}\!\left(\!D\Psi\!\right)^2 \right)\!-\!{1 \over 2}
e^{-2\left(\Phi+U\right)}(\!\partial_0\!\Psi\!)^2\right)\ .
\label{acp4}
\end{equation}
The last term of (\ref{acp4}) can be written
\begin{equation}
e^{-2 \left(\Phi+U\right)} (\partial_0\Psi)^2 = \partial_0
\left(e^{-2 \left(\Phi+U\right)} \Psi \partial_0\Psi \right) -
e^{-2 \left(\Phi+U\right)} \partial_0 \partial_0 \Psi \ . \label{axvan}
\end{equation}
The first term on the rhs of (\ref{axvan}) is a vanishing
surface term and the second term vanishes by (\ref{axderiv}).
With all dependence on $x^0$ explicitly removed, the action
(\ref{acp4})
can be rewritten as an effective three-dimensional action on $\Sigma$,
\begin{equation}
I=\int_{\Sigma} d^3 \! x \sqrt{\gamma} \left(\!R(\gamma) -
{1 \over 2} \!\left(DU \right)^2 - {1 \over 4} e^{2U} V_{ij} V^{ij} -
{1 \over 2} \! \left( \left(D \Phi \right)^2 \!-\! e^{- 2 \Phi}
\!\left(D \Psi \right)^2 \right) \right) \ ,
\label{aca3}
\end{equation}
where all indices are now raised and lowered with $\gamma$.
The Hodge dual of $V_{ij}$ is $V_i$
given by
\begin{equation}
V_{ij} = {\epsilon_{ij}}^k V_k \ , \label{start}
\end{equation}
where $V_i$ obeys the conservation equation
\begin{equation}
D_i V^i = 0 \ , \label{cont}
\end{equation}
As before, a new action is defined by adding the constraint
(\ref{cont}), multiplied by a Lagrange multiplier V, to (\ref{aca3})
\begin{equation}
I=\int_{\Sigma} d^3 \! x \sqrt{\gamma} \left(\!R(\gamma)\! -
\!{1 \over 2} \!\left(\!DU \right)^2 \!-\! {1 \over 2} e^{2U} V_i V^i
\!+\!
V\!D_i V^i \!-\! {1 \over 2} \!\left( \left(\!D \Phi \!\right)^2 \!-\!
e^{- 2 \Phi}\!\left(\!D \Psi\!\right)^2 \right) \right) \ .
\label{acl3}
\end{equation}
The equation of motion for $V_i$ is
\begin{equation}
e^{2U} V_i = - D_i V \ . \label{eomt}
\end{equation}
Substituting (\ref{eomt}) into (\ref{acl3}), integrating by parts, and
removing the surface term gives the final form of the action,
\begin{equation}
I=\int_{\Sigma} d^3 \! x \sqrt{\gamma} \left(\!R(\gamma) - {1 \over 2}
\!\left( \left(DU \right)^2 \!-\! e^{-2U} \!\left(DV \right)^2 \right)
- {1 \over 2} \!\left( \left(D \Phi \right)^2 \!-\! e^{- 2 \Phi}
\!\left(D \Psi \right)^2 \right) \right) \ . \label{ac3}
\end{equation}
This action describes a three dimensional spacetime with metric
$\gamma_{ij}$, scalars $U$ and $\Phi$, and pseudo-scalars $V$ and
$\Psi$.
There now appears to be two $O(2,1)$ non-linear $\sigma$-models, each
with
its own $SL(2, R)$ isometry: the original S-duality and a new isometry
which is actually the $SL(2, R)$ Ehlers' group of symmetries from
dimensionally reduced General Relativity \cite{gary}. So there is a
new
duality of the four dimensional string effective action with one
killing
symmetry, and we have termed this E-duality.  Obviously the
symmetries of this action are larger than the combined S-duality and
$O(1,1)$ target space duality: two independent $SL(2, R)$ symmetries
given by (\ref{dilat})-(\ref{mix}) and the same transformations
with $U$, $V$ replacing $\Phi$, $\Psi$ respectively; and (\ref{ac3})
is also invariant under
\begin{eqnarray}
U & \leftrightarrow & \Phi \nonumber \\
V & \leftrightarrow & \Psi \label{inv} \ .
\end{eqnarray}
This final transformation is highly suggestive of discrete target
space duality. The total symmetry here is
\begin{equation}
SL(2, R) \otimes SL(2, R) \simeq O(2,2; R) \ ,
\end{equation}
though of course this $O(2,2)$ symmetry is totally distinct to the
T-duality $O(2,2)$ symmetry when there are two isometries.
The task is now to find an explicit realization in terms of field
redefinitions
of the $O(1,1)$  target space duality under which (\ref{ac3}) should
be invariant,
and to see how this mixes with the duality symmetries found above.

{\center \section{\bf T-duality of the effective action}}

Unlike the case of S-duality, the dimension of the target space is not
critical
to T-duality; the only requirement is the presence of an isometry. For
this
reason the following analysis will be performed in arbitrary dimension,

specializing to four dimensions later. We return to the
original string frame action (\ref{acs4}), now generalized to
D-dimensions.
Once again, we assume the existence of a one-parameter isometry
generated by
the Killing vector $K$, and now $\Sigma$ is the $d=D-1$ dimensional
base manifold.
The procedure in (\ref{lie})-(\ref{R}) follows and the string frame
Ricci scalar is
\begin{equation}
\tilde{R}(\tilde{g})=e^U\tilde{R}(\tilde{\gamma}) + (d-2)e^U D^2U -
{1 \over 4} (d^2-5d+8)e^U(DU)^2 - {1 \over 4} e^{3U}V_{ij}V^{ij}
\ , \label{RD}
\end{equation}
and the determinant of $\tilde{g}_{\mu\nu}$ is
\begin{equation}
\det (\tilde{g}_{\mu\nu}) = e^{U(1-d)}
\det (\tilde{\gamma}_{ij}) \ . \label{detD}
\end{equation}
The dilaton term is given again by (\ref{dilred}), but
the reduction of the Kalb-Ramond field is more involved.  Using the
projection operator (\ref{proj}), $B_{\mu\nu}$ is decomposed into
\begin{eqnarray}
b = & K^\mu K^\nu B_{\mu\nu} & = \ 0 \nonumber \\
b_i = & K^\mu h^\nu _{\ i} B_{\mu\nu} & = \ B_{0i} \\
b_{ij} = & h^\mu _{\ i} h^\nu _{\ j} B_{\mu\nu} & = \ B_{ij}-\omega_i
B_{0j} - \omega_j B_{i0} \ , \nonumber
\end{eqnarray}
and can be written in the form
\begin{equation}
B_{\mu\nu}=\left (\begin{array}{lc} 0 & b_j \\[9pt] -b_i &
b_{ij}+\omega_{[i}b_{j]} \end{array} \right) \ .
\end{equation}
An analogue of $V_{ij}$ can be constructed for $b_i$ ,
\begin{equation}
W_{ij} = \partial_i b_j - \partial_j b_i \ .
\end{equation}
Now $H_{\mu\nu\rho}$ is decomposed into
\begin{eqnarray}
H_{0ij} & = & -W_{ij} \\
H_{ijk} & = & \partial_{[i} b_{jk]} + b_{[i}V_{jk]}
-\omega_{[i}W_{jk]} \label{Hijk}
\end{eqnarray}
and the $H$ field term in (\ref{acs4}) is written
\begin{equation}
H_{\mu\nu\rho} H^{\mu\nu\rho} = 3 e^U W_{ij} W^{ij} + e^{3U}
\theta_{ijk} \theta^{ijk} \ , \label{theta}
\end{equation}
where
\begin{equation}
\theta_{ijk} = \partial_{[i}b_{jk]} + b_{[i}V_{jk]} \ .
\label{thetadef}
\end{equation}

\noindent Inserting (\ref{det}), (\ref{R}), (\ref{dilred}), and
(\ref{theta}) in (\ref{acs4}),
the action becomes independent of $x^0$ and reduces to
\begin{eqnarray}
\lefteqn{I=\int_\Sigma d^d\! x \sqrt{\tilde{\gamma}} e^{\Phi+{1 \over
2} U(3-d)} \left(\tilde{R}(\tilde{\gamma})+ (d\!-\!2)D^2U -
{1 \over 4} (d^2\!\! -\! 5d\! +\! 8)(\!DU)^2 +
(\!D \Phi)^2 \right.} \hspace{150pt} \nonumber \\ \label{isd} \\ & &
\hspace{20pt} \left.\ \ -{1\over4}e^{2U}V_{ij} V^{ij} - {1 \over 4}
W_{ij} W^{ij} - {1 \over 12} e^{2U} \theta_{ijk} \theta^{ijk} \right)
\ . \nonumber
\end{eqnarray}
Following \cite{Busch}, (\ref{isd}) should be invariant under the
$O(1,1)$ duality symmetry, consisting of a one-parameter scaling
\begin{eqnarray}
\tilde{g}_{00} & \rightarrow & e^a \tilde{g}_{00} \ , \nonumber \\
\tilde{g}_{0i} & \rightarrow & e^{a/2} \tilde{g}_{0i} \ , \\
B_{0i} & \rightarrow & e^{a/2} B_{0i} \ , \nonumber \label{scale}
\end{eqnarray}
and a discrete inversion
\begin{eqnarray}
\hspace{112pt}\tilde{g}_{00} &\rightarrow& 1/\tilde{g}_{00} \ ,
\nonumber
\\ \tilde{g}_{0\alpha} &\rightarrow& B_{0\alpha}/\tilde{g}_{00}
\ , \nonumber \\ \tilde{g}_{\alpha\beta} &\rightarrow&
\tilde{g}_{\alpha\beta} -
(\tilde{g}_{0\beta} \tilde{g}_{0\alpha} - B_{0\alpha} B_{0\beta} )/
\tilde{g}_{00} \label{gt}  \ ,  \\
 B_{0\alpha} &\rightarrow& \tilde{g}_{0\alpha}/\tilde{g}_{00} \ ,
\nonumber
\\ B_{\alpha\beta} &\rightarrow& B_{\alpha\beta} - \tilde{g}_{0[\beta}
B_{\alpha]0}/\tilde{g}_{00}
\ . \nonumber \label{bt}
\end{eqnarray}
A dilaton transformation is required in both these transformations to
ensure
conformal invariance is preserved in the dual theory (at one loop).
This is
\begin{eqnarray}
\hspace{20pt}\phi &\rightarrow& \phi + \ln \tilde{g}_{00} \ .
\label{dila}
\end{eqnarray}
The explicit form of the symmetry can be rewritten in terms of
redefinitions of the fields appearing in the action (\ref{isd}).
The scaling can be realized in the following field redefinitions:
\begin{eqnarray}
U & \rightarrow & U + a \nonumber \\
\omega_i & \rightarrow & e^{-a/2} \omega_i \nonumber \\
b_i & \rightarrow & e^{a/2} b_i \label{o1s} \\
\tilde{\gamma}_{ij} & \rightarrow & e^a \tilde{\gamma}_{ij} \nonumber
\\ \Phi & \rightarrow & \Phi - a/2 \nonumber \ ,
\end{eqnarray}
under which the target space fields become
\begin{equation}
\tilde{g}_{\mu\nu} = \left (\begin{array}{lc} e^{U+a} &
e^{U+a/2}\omega_j \\[9pt]
e^{U+a/2}\omega_i &  e^U\omega_i \omega_j + e^{-U}\tilde{\gamma}_{ij}
\end{array}
\right) \ , \hspace{30pt} B_{\mu\nu}=\left (\begin{array}{lc} 0 &
e^{a/2} b_j \\
[9pt] - e^{a/2} b_i & b_{ij}+\omega_{[i}b_{j]} \end{array} \right) \ .
\end{equation}
The inversion part of the symmetry is similarly realised as
\begin{eqnarray}
\hspace{45pt} U & \rightarrow & -U \nonumber \\
\omega_i & \rightarrow & b_i \nonumber \\
b_i & \rightarrow & \omega_i \label{o1i} \\
\tilde\gamma_{ij} & \rightarrow & e^{-2U} \tilde{\gamma}_{ij}
\nonumber \\ b_{ij} & \rightarrow & b_{ij} + \omega_ib_j -
\omega_jb_i \nonumber \\ \Phi & \rightarrow & \Phi + U \nonumber \ ,
\end{eqnarray}
under which the target space fields become
\begin{equation}
\tilde{g}_{\mu\nu}=\left(\begin{array}{lc} e^{-U} & e^{-U}b_j \\[9pt]
e^{-U}b_i &  e^{-U} b_i b_j + e^{-U}\tilde{\gamma}_{ij} \end{array}
\right)\ ,\hspace{30pt}B_{\mu\nu}=\left(\begin{array}{lc} 0 & \omega_j
\\[9pt] -\omega_i & b_{ij} \end{array} \right) \ .
\end{equation}

\noindent It can be shown that (\ref{isd}) is indeed invariant under
field
redefinitions (\ref{o1s}) and (\ref{o1i}).  Furthermore, the
d-dimensional
metric $e^{-U}\tilde{\gamma}_{ij}$, and the d-dimensional 3-form
$\theta_{ijk}$
are both invariant under these redefinitions.

\noindent We can see from the field redefinitions in (\ref{o1i}) that
the discrete
duality symmetry involves exchanging the $V_{ij}$ and $W_{ij}$ in
(\ref{isd}).
It would be interesting to see whether (\ref{isd}) could be put in a
more appropriate form to examine this field exchange.  First the
$D^2U$ term is removed by integrating by parts then removing the
total divergence, giving
\begin{eqnarray}
\lefteqn{I=\int_\Sigma d^d\! x \sqrt{\tilde{\gamma}} e^{\Phi+{1 \over
2}U(3-d)} \left(\tilde{R}(\tilde{\gamma}) +
{1 \over 4}(d\!-\!1)(d\!-\!4)(\!DU)^2 - (d\!-\!2)DU\,D\Phi +
(\!D\Phi\!)^2 \right.} \hspace{150pt} \nonumber \\ \label{isd2}
\\ & & \hspace{20pt} \left.\ \  - {1 \over 4}e^{2U}V_{ij} V^{ij} -
{1 \over 4} W_{ij} W^{ij} -
{1 \over 12} e^{2U} \theta_{ijk} \theta^{ijk} \right) \ . \nonumber
\end{eqnarray}
This action can be transformed to the Einstein frame with respect to
$\tilde{R}(\tilde{\gamma})$ by
introducing a new metric $\gamma_{ij}$ where
\begin{equation}
\gamma_{ij} = \exp\left({\frac{2\Phi + U(3\!-\!d)}{d\!-\!2}}\right)
\tilde{\gamma}_{ij} \label{gam}
\end{equation}
and thus (\ref{isd2}) becomes (for $d\neq 2$)
\begin{eqnarray}
\lefteqn{I=\int_\Sigma d^d\! x \sqrt{\gamma}
\left(R(\gamma) - {1 \over 4}{(d\!-\!1) \over (d\!-\!2)}(\!DU)^2 -
{1 \over (d\!-\!2)}DU\,D\Phi - {1 \over (d\!-\!2)}(\!D\Phi\!)^2
\right.}
\hspace{20pt} \nonumber \\ \nonumber \\ & & \left.\ \ - {1 \over 4}
\exp{\left({\frac{2\Phi\!+\!U(d\!-\!1)}{d\!-\!2}}\right)} V_{ij}
V^{ij} -
{1 \over 4}\exp{ \left({\frac{2\Phi\!+\!U(3\!-\!d)}{d\!-\!2}}\right)}
W_{ij} W^{ij} \right. \label{isd3} \\ \nonumber \\ & & \hspace{190pt}
\left. -{1 \over 12} \exp{\left({\frac{4\Phi\!+\!2U}{d\!-\!2}}\right)}
\theta_{ijk} \theta^{ijk} \right) \ . \nonumber
\end{eqnarray}
The $U$ and $\Phi$ terms can be decoupled by defining new variables
$\tilde{U}$ and $\tilde{\Phi}$ given by
\begin{equation}
U=2(1\!+\!\sqrt{d\!-\!2})\,\tilde{U}\ -\ 2(1\!+\!\sqrt{d\!-\!2})\,
\tilde{\Phi}\ ,\label{U-}
\end{equation}
\begin{equation}
\Phi=(d\!-\!3)\,\tilde{U}\,+\,(d\!-\!1\!+\!2\sqrt{d\!-\!2})\,
\tilde{\Phi}\ .\label{phi-}
\end{equation}
\noindent Inserting (\ref{U-}) and (\ref{phi-}) into (\ref{isd3})
gives the most symmetric form of the action,
\begin{eqnarray}
I=\int_\Sigma d^d\! x \sqrt{\gamma} \left(R(\gamma)
- c(\!D\tilde{U})^2 - c(\!D\tilde{\Phi}\!)^2\right. \hspace{200pt}
\nonumber \\ \hspace{100pt} \left.-{1 \over 4}e^{x\tilde{U}+
y\tilde{\Phi}} V_{ij} V^{ij}-{1 \over 4} e^{x\tilde{\Phi}+y\tilde{U}}
W_{ij} W^{ij}-{1 \over 12} e^{(\!x+y\!)(\tilde{U}+\tilde{\Phi}\!)}
\theta_{ijk} \theta^{ijk} \right) \label{acd}
\end{eqnarray}
\noindent where
\begin{eqnarray}
c = 2 \left(d\!-\!1+2\sqrt{d\!-\!2}\right) \hspace{40pt} x =
4+2{d\!-\!1 \over {\sqrt{d\!-\!2}}}  \hspace{40pt} y = 2{3\!-\!d
\over {\sqrt{d\!-\!2}}}  \ . \label{cxy}
\end{eqnarray}
\noindent The $O(1,1)$ duality transformations (\ref{o1s}),
(\ref{o1i}) can be rewritten in terms of the new variables defined in
(\ref{gam}), (\ref{U-}), and (\ref{phi-}). The scaling part becomes
\begin{eqnarray}
\tilde{U} & \rightarrow & \tilde{U} + a \nonumber \\
\tilde{\Phi} & \rightarrow & \tilde{\Phi} - a \nonumber \\
V_{ij} & \rightarrow & e^{a(y-x)/2} V_{ij} \label{O1s} \\
W_{ij} & \rightarrow & e^{a(x-y)/2} W_{ij} \nonumber \\
\theta_{ijk} & \rightarrow & \theta_{ijk} \nonumber \\
\gamma_{ij} & \rightarrow & \gamma_{ij} \nonumber \ ,
\end{eqnarray}
and the inversion,
\begin{eqnarray}
\tilde{U} & \leftrightarrow & \tilde{\Phi} \hspace{50pt} \nonumber \\
V_{ij} & \leftrightarrow & W_{ij} \label{O1i} \\
\theta_{ijk} & \rightarrow & \theta_{ijk} \nonumber \\
\gamma_{ij} & \rightarrow & \gamma_{ij} \nonumber \ .
\end{eqnarray}
The discrete inversion of the O(1,1) transformation is now explicitly
shown in terms of a simple field exchange symmetry.

\noindent Returning to the case of $d=3$ ($D=4$), the action
(\ref{acd}), after
rescaling $\tilde{\Phi}$ and $\tilde{U}$ by a factor of four, becomes
\begin{eqnarray}
I=\int_\Sigma d^3\! x \sqrt{\gamma} \left(R(\gamma)\!
-\!{1 \over 2}(D\tilde{U})^2\! -\! {1 \over 2}(D\tilde{\Phi})^2\!-\!{1
\over 4}e^{2\tilde{U}} V_{ij} V^{ij}\!-\!{1 \over 4}
e^{2\tilde{\Phi}} W_{ij} W^{ij}\!   -\!{1 \over 12}
e^{2\tilde{U}}e^{2\tilde{\Phi}} \theta_{ijk} \theta^{ijk} \right)
\label{act3} \ .
\end{eqnarray}
The 3-form $\theta_{ijk}$ in three dimensions must be some scalar,
$\theta$, multiplied by the volume form,
\begin{equation}
\theta_{ijk} = \theta \epsilon_{ijk} \ .
\end{equation}
Thus, an equivalent action to (\ref{act3}) would be that action with
the $\theta_{ijk} \theta^{ijk}$ term replaced by
\begin{equation}
{\cal{L}}_\theta = -{1 \over 2} e^{2\tilde{U}}e^{2\tilde{\Phi}}
\theta^2 + {1 \over 6} \Lambda\epsilon_{ijk}
\left(\theta\epsilon^{ijk} -
\partial^{[i}b^{jk]} - b^{[i}V^{jk]} \right) \ , \label{theta2}
\end{equation}
where the equation of motion for the Lagrange multiplier, $\Lambda$,
gives back the definition of $\theta_{ijk}$ as a constraint
(\ref{thetadef}).
$\theta$ can be consistently integrated out of (\ref{theta2}) via its
equation of motion,
\begin{equation}
e^{2\tilde{U}}e^{2\tilde{\Phi}}\theta = \Lambda \ ,
\end{equation}
to give the first order form for $\cal{L}_{\theta}$,
\begin{equation}
{\cal{L}}_{\theta} = {1 \over 2} e^{-2(\tilde{U}+\tilde{\Phi})}
\Lambda^2 -
{1 \over 6}\Lambda\epsilon_{ijk} \left(\partial^{[i}b^{jk]} +
b^{[i}V^{jk]}\right) \ . \label{theta3}
\end{equation}
The equation of motion for $b_{ij}$,
\begin{equation}
\partial_i \Lambda = 0 \ ,
\end{equation}
reveals that $\Lambda$ is in fact a constant, not a scalar \cite{cosm},
and the equation of motion for $\Lambda$,
\begin{equation}
e^{-2(\tilde{U}+\tilde{\Phi})} \Lambda = {1 \over 6}\epsilon_{ijk}
\left(\partial^{[i}b^{jk]} + b^{[i}V^{jk]} \right)
\ , \label{lambda}
\end{equation}
enables us to write
\begin{equation}
{\cal{L}}_{\theta} = -{1 \over 2} e^{-2(\tilde{U}+\tilde{\Phi})}
\Lambda^2 \ .
\end{equation}
However, using (\ref{Hijk}) in (\ref{lambda}) gives
\begin{equation}
e^{-2(\tilde{U}+\tilde{\Phi})} \Lambda = {1 \over 6}
e^{-3\tilde{U}}\epsilon_{ijk} H^{ijk} \ , \label{lambdaH}
\end{equation}
and recalling the relations (\ref{hogH}), (\ref{eomH}), and
(\ref{ax3}) we obtain
\begin{equation}
{\cal{L}}_{\theta} = -{1 \over 2}
e^{-2(\tilde{U}+\tilde{\Phi})} (\partial_0 \Psi)^2 \ .
\end{equation}
As before, integrating this by parts (\ref{axvan}) and using
(\ref{axderiv}) shows that
${\cal{L}}_\theta$ vanishes. In other words, the cosmological term
that should arise when there is a quadratic n-form term in n
dimensions \cite{cosm} actually vanishes in this particular case.
The 2-forms $V_{ij}$ and $W_{ij}$ may be dualized using the procedure
set out in (\ref{start})-(\ref{ac3}) into the pseudo-scalars $V$ and
$\Psi$ respectively and the final form of the action becomes
\begin{equation}
I=\int_{\Sigma} d^3 \! x \sqrt{\gamma} \left(R(\gamma) -
{1 \over 2} \left( \left(D\tilde{U} \right)^2 - e^{-2\tilde{U}}
\left(DV \right)^2 \right)
- {1 \over 2}  \left( \left(D \tilde{\Phi} \right)^2 - e^{- 2
\tilde{\Phi}} \left(D \Psi \right)^2 \right) \right) \ , \label{a3}
\end{equation}
which is identical to (\ref{ac3}). The explicit action of $O(1,1)$
duality upon (\ref{ac3}) is now known and is given by the following
field redefinitions
\begin{eqnarray}
\tilde{U} & \rightarrow & \tilde{U} + a \nonumber \\
\tilde{\Phi} & \rightarrow & \tilde{\Phi} - a \nonumber \\
V & \rightarrow & e^{-a} V \label{O11s} \\
\Psi & \rightarrow & e^a \Psi \nonumber \ ,
\end{eqnarray}
and
\begin{eqnarray}
\tilde{U} & \leftrightarrow & \tilde{\Phi} \hspace{25pt} \nonumber \\
V & \leftrightarrow & \Psi \label{O11i} \ .
\end{eqnarray}
The scaling of $U$ and $V$ is already part of the E-duality
transformations,
and the scaling of $\Phi$ and $\Psi$ is already part of the S-duality
transformations. It appears that $O(1,1)$ duality simply interchanges
the two $SL(2, R)$ symmetries. Thus, there is a very extensive
symmetry of the four
dimensional bosonic effective action with one isometry, which can be
represented by
{\bf
\begin{center}
\begin{tabular}{ccc}
E-duality & \quad $\longleftarrow\!\!\longrightarrow$ \quad &\,
S-duality \\ $SL(2, R)$  & \quad $O(1,1)$ \quad & \, $SL(2,  R)$ \\
& \,T-duality &
\end{tabular}
\end{center}}
\noindent and the total group structure is thus $O(2,2; R)$.
In \cite{Sen3d} it has been shown that heterotic string theory
compactified on a
$7$-torus has a duality group of $O(8,24; Z)$ which consists
of $O(7,23; Z)$
T-duality and $SL(2, Z)$ S-duality. This is a generalization
of the above result.
{\center \section{More general bosonic actions}}

Bosonic string theory suffers from a tachyonic ground
state. Fortunately,
in the construction of more realistic string theories, this state is
projected out. In the above, we have assumed that the tachyon field
has been set to zero, which is consistent with treating the action
as a truncation of a supersymmetric theory. However,
for completeness, we will discuss the case of a non-trivial tachyon.
The modified $4$-d effective action, generalizing (\ref{acs4}), is
\begin{equation}
I=\int_{\cal{M}} d^4 \! x \sqrt{\tilde{g}} e^{\Phi} \left(\tilde{R}
(\tilde{g}) + \left(\nabla
\Phi \right)^2 - {1 \over 12} H_{\mu\nu\rho} H^{\mu\nu\rho} -
\left(\nabla T \right)^2 + V(T) \right)\ , \label{acs4T}
\end{equation}
where $T$ is the tachyon and $V(T)$ is the tachyon potential
\cite{tach}. Transforming (\ref{acs4T}) to the form in which
S-duality is manifest, we find
\begin{equation}
I=\int_{\cal{M}} d^4 \! x \sqrt{g} \left(R(g) - {1 \over
2}  \left\{ \left(\nabla \Phi \right)^2 - e^{- 2 \Phi} \left(\nabla
\Psi \right)^2 \right\} - \left(\nabla T \right)^2 + e^{-\Phi} V(T)
\right) \ . \label{aca4T}
\end{equation}
Thus, while the inclusion of the tachyon kinetic term is trivially
compatible with S-duality, the potential cannot transform so as to
keep the
last term of (\ref{aca4T}) invariant. Thus, the tachyon potential
explicitly breaks S-duality. Proceeding as before and reducing
(\ref{aca4T}) to an effective $3$-d action gives
\begin{equation}
\!\!\!\!I\!=\!\int_{\Sigma}\! d^3 \! x \!\sqrt{\gamma}
\left(\!R(\gamma)
\!-\! {1 \over 2} \!\left( \left(\!DU \right)^2 \!-\! e^{-2U}\!
\!\left(\!DV \right)^2 \right) \!-\! {1 \over 2} \!\left(
\left(\!D \Phi \!\right)^2 \!-\! e^{- 2 \Phi} \!\left(\!D \Psi \!
\right)^2 \right)\!-\! \left(\!D T \right)\!^2 \!+\! e^{-(U\!+\Phi)}
V(T) \!\right) \label{ac3T}
\end{equation}
So the potential also explicitly breaks E-duality. However,
(\ref{ac3T})
is invariant under the T-duality scaling and inversion symmetries
(\ref{O11s}) and (\ref{O11i}), with $T$ and $V(T)$ remaining
unchanged.
This is to be expected as including the tachyon term in the worldsheet
action \cite{tach} does not affect the worldsheet duality
transformation
\cite{Busch}. In arbitrary dimension, the effective action is
\begin{eqnarray}
I=\int_\Sigma d^d\! x \sqrt{\gamma} \left(R(\gamma)
- c(\!D\tilde{U})^2 - c(\!D\tilde{\Phi}\!)^2 - \left(D T \right)^2 +
e^{z(U+\Phi)} V(T)\right. \hspace{130pt}\nonumber \\
\hspace{100pt} \left.-{1 \over 4}e^{x\tilde{U}+
y\tilde{\Phi}} V_{ij} V^{ij}-{1 \over 4} e^{x\tilde{\Phi}+y\tilde{U}}
W_{ij} W^{ij}-{1 \over 12} e^{(\!x+y\!)(\tilde{U}+\tilde{\Phi}\!)}
\theta_{ijk} \theta^{ijk} \right) \label{acdT}
\end{eqnarray}
with $c$, $x$, $y$ given by (\ref{cxy}) and
\begin{equation}
z=4-2d-2{\sqrt{d-2}} \ .
\end{equation}

A cosmological constant term, $\Lambda$, may be added to the original
action
(\ref{acs4}) to give a $c=0$ string theory with a target space of
dimension less than the critical dimension \cite{beta},
\begin{equation}
I=\int_{\cal{M}} d^d \! x \sqrt{\tilde{g}} e^{\Phi} \left(\tilde{R}
(\tilde{g}) + \left(\nabla \Phi \right)^2 - {1 \over 12}
H_{\mu\nu\rho} H^{\mu\nu\rho} + \Lambda \right) \ , \label{acs4C}
\end{equation}
where
\begin{equation}
\Lambda = {{2(d-26)} \over {3\alpha'}} \ .
\end{equation}
This term enters the action in the same way as the tachyon potential,
and in the precisely the same manner explicitly breaks S and
E-duality. It is consistent with $O(1,1)$ T-duality.


{\center \section{Duality in $d>4$}}

\noindent Returning to the original d-dimensional action (\ref{isd}),
there is another one-parameter isometry that does not belong to the
expected $O(1,1)$ group of T-duality transformations:
\begin{eqnarray}
U & \rightarrow & U + a \nonumber \\
\Phi & \rightarrow & \Phi - a(3\!-\!d)/2 \label{ps1}\\
\omega_i & \rightarrow & e^{-a} \omega_i \nonumber \\
b_{ij} & \rightarrow & e^{-a} b_{ij} \nonumber \ ,
\end{eqnarray}
and $\tilde{g}_{\mu\nu}$, $B_{\mu\nu}$ become
\begin{equation}
\tilde{g}_{\mu\nu} = \left (\begin{array}{lc} e^{U+a} & e^{U}\omega_j
\\[9pt] e^{U}\omega_i &  e^{U-a}\omega_i \omega_j + e^{-(U+a)}
\tilde{\gamma}_{ij} \end{array} \right) \ , \hspace{30pt}
\tilde{B}_{\mu\nu}=\left (\begin{array}{lc} 0 & b_j
\\[9pt] -b_i & e^{-a}(b_{ij}+\omega_{[i}b_{j]}) \end{array} \right) \ .
\end{equation}
Note that neither $e^{-U}\tilde{\gamma}_{ij}$ nor $\theta_{ijk}$ are
invariant under (\ref{ps1}). When examined, this new symmetry combines
with the standard scaling symmetry of $O(1,1)$ duality given in
(\ref{o1s}) to give a much more general set of scalings,
\begin{eqnarray}
\tilde{U} & \rightarrow & \tilde{U} + a \nonumber \\
\tilde{\Phi} & \rightarrow & \tilde{\Phi} + b \nonumber \\
V_{ij} & \rightarrow & e^{-(ax+by)/2} V_{ij} \label{pps} \\
W_{ij} & \rightarrow & e^{-(ay+bx)/2} W_{ij} \nonumber \\
b_{ij} & \rightarrow & e^{-(a+b)(x+y)/2}b_{ij} \nonumber \\
\gamma_{ij} & \rightarrow & \gamma_{ij} \nonumber \ ,
\end{eqnarray}
of which (\ref{O1s}) is the special case of $a=-b$.
Thus, the symmetry group of the bosonic effective action with one
Killing vector is larger than the target space duality $O(1,1)$
symmetry.
At first sight the extra symmetry seems to be a remnant of the two
independent $SL(2, R)$ symmetries that exist in four dimensions.
If the basic T-duality scaling is removed from (\ref{pps}) then the
remaining scaling can be written as a transformation on the original
fields of (\ref{acs4})
\begin{eqnarray}
\tilde{g}_{\mu\nu} & \rightarrow & e^a \tilde{g}_{\mu\nu} \nonumber \\
B_{\mu\nu} & \rightarrow & e^a B_{\mu\nu} \label{pp} \\
\Phi & \rightarrow & \Phi + a(1-d)/2 \nonumber \ .
\end{eqnarray}
The overall effect of (\ref{pp}) is to rescale the string coupling
constant, $e^{\Phi}$. At the level of the worldsheet, this amounts
to adding a topological term to the worldsheet action. Note that
this extra symmetry is respected by neither the tachyon potential
(\ref{acdT}) nor the cosmological constant term of the previous
section.
{\center\section{\bf Conclusion}}

The combined effects of S and T-duality have been investigated in the
context of four dimensional closed bosonic string theory with one
compact dimension, where they have been found to form an extended
$O(2,2)$ duality group, incorporating the Ehlers' symmetry of General
Relativity. We have found the most symmetric form
of the effective action which clearly demonstrates its manifest
invariance under S and T-duality, and under the `new' E-duality.

The interchange between S and E-duality is particularly interesting
as it shows that  with a one dimensional compactification, the
effective non-gravitational fields are in a one-to-one correspondence
with the Kaluza-Klein fields derived from the dimensionally reduced
metric. The dilaton appears on much the same footing as the modulus
field $U$, despite its very different appearance in the bosonic
worldsheet action. It would be interesting to see how this picture
develops in the supersymmetric theories and exactly what structures
are interchanged there under T-duality.

The inclusion of the tachyon field does not affect T-duality of the
effective action, which is to be expected as T-duality is an exact
symmetry of bosonic string theory \cite{gpr}. However, the tachyon
potential explicitly breaks S-duality. It would appear that although
S-duality is a symmetry of the massless sector of bosonic string
theory, it cannot be an exact symmetry. It remains to be seen
whether S-duality will survive as an exact symmetry of the
supersymmetric string theories.

A further symmetry has been discovered which demonstrates the
invariance of the bosonic effective action under a rescaling of
the string coupling constant.
\newpage

After the completion of this work,  it was noticed that \cite{bakas2}
contains
some overlap with the results presented here. I would like to thank my
supervisor, Malcolm Perry, for thoughts and guidance during the course
of this work. I am also grateful to Lloyd Alty, Gary Gibbons,
Neil Lambert, Simon Ross, Edward Teo and Paul Townsend for useful
discussions and comments.


\begin{thebibliography}{xxx}
\bibitem{MO} C.Montonen and D.Olive, Phys. Lett. B72 (1977) 117.
\bibitem{Sen} A. Sen, Int. J. Mod. Phys. A9 (1994) 3707.
\bibitem{gpr} A. Giveon, M. Porrati, and E. Rabinovici,
   Phys. Rep. 244 (1994) 77, and references therein.
\bibitem{bakas1} I. Bakas, Nucl. Phys. B428 (1994) 374.
\bibitem{beta} C. Callan, D. Friedan, E. Martinec and M.J.
   Perry, Nucl. Phys. B262 (1985) 593.
\bibitem{gary} G.W. Gibbons and S.W. Hawking, Commun. Math. Phys.
   66 (1979) 291.
\bibitem{Busch} T.H. Buscher, Phys. Lett. B194 (1987) 59.
\bibitem{cosm} A. Aurilia, H. Nicolai, and P.K. Townsend,
   Nucl. Phys. B176 (1980) 509.
\bibitem{Sen3d} A. Sen, Nucl. Phys. B434 (1995) 179.
\bibitem{tach} V.A. Kostelecky and M.J. Perry,
   Nucl. Phys. B414 (1994) 174.
\bibitem{bakas2} I. Bakas, ``Spacetime interpretation of S-duality
   and supersymmetry violations of T-duality'', preprint
   hep-th/9410104.
\end{thebibliography}
\end{document}